\documentclass[
 reprint,
 amsmath,amssymb,
 aps,showkeys
]{revtex4-2}

\usepackage{graphicx}
\usepackage{bm}
\usepackage{dcolumn}
\usepackage{bm}
\usepackage[whole]{bxcjkjatype}
\usepackage{cleveref}
\usepackage{empheq}
\usepackage{enumerate}
\usepackage{comment}
\usepackage{algorithm}
\usepackage{algpseudocode}
\usepackage{physics2}
\usepackage[hang,small,bf]{caption}
\usepackage[subrefformat=parens]{subcaption}
\usephysicsmodule{ab, braket, doubleprod, xmat}

\usepackage{xcolor}

\crefname{equation}{Eq.}{Eq.}
\crefname{figure}{Fig.}{Fig.}
\crefname{table}{Table}{Table}

\usepackage{array}
\newcolumntype{M}[1]{>{\centering\arraybackslash}m{#1}}
\usepackage{colortbl}

\newcommand{\argmin}{\mathop{\rm argmin}\limits}

\newcommand{\Tr}{\operatorname{Tr}}

\crefname{algorithm}{Algorithm}{Algorithms}

\crefname{dfn}{Definition}{Definitions}

\crefname{prop}{Proposition}{Propositions}

\crefname{thm}{Theorem}{Theorems}

\algnewcommand\algorithmicinput{{\bfseries\gtfamily Input}}%
\algnewcommand\algorithmicoutput{{\bfseries\gtfamily Output}}%
\algnewcommand\AlgInput{\item[\algorithmicinput]}%
\algnewcommand\AlgOutput{\item[\algorithmicoutput]}%
\algrenewcommand\Return{\State\textbf{return} }%

\usepackage{xcolor}
\begin{document}
\newcommand{\mh}[1]{{\textcolor{cyan}{\bf #1}}}

\preprint{APS/123-QED}

\title{Structured quantum learning via em algorithm for Boltzmann machines}

\author{Takeshi Kimura${}^1$}
 \email{tkimura@nagoya-u.jp}
\author{Kohtaro Kato${}^1$}
 \email{kokato@i.nagoya-u.ac.jp}
\author{Masahito Hayashi${}^{2, 3, 4}$}
 \email{hmasahito@cuhk.edu.cn (Corresponding author)}
\affiliation{
 ${}^1$Department of Mathematical Informatics, Graduate School of Informatics, Nagoya University, Furo-cho, Chikusa-ku, Nagoya 464-8601, Japan
}
\affiliation{
 ${}^2$School of Data Science, The Chinese University of Hong Kong, Shenzhen, Longgang District, Shenzhen, 518172, China
}
\affiliation{
 ${}^3$International Quantum Academy, Futian District, Shenzhen 518048, China
}
\affiliation{
 ${}^4$Graduate School of Mathematics, Nagoya University, Furo-cho, Chikusa-ku, Nagoya, 464-8601, Japan
}

\begin{abstract}
Quantum Boltzmann machines (QBMs) are generative models with potential advantages in quantum machine learning, yet their training is fundamentally limited by the barren plateau problem, where gradients vanish exponentially with system size. We introduce a quantum version of the em algorithm, an information-geometric generalization of the classical Expectation-Maximization method, which circumvents gradient-based optimization on non-convex functions. Implemented on a semi-quantum restricted Boltzmann machine (sqRBM)—a hybrid architecture with quantum effects confined to the hidden layer—our method achieves stable learning and outperforms gradient descent on multiple benchmark datasets. These results establish a structured and scalable alternative to gradient-based training in QML, offering a pathway to mitigate barren plateaus and enhance quantum generative modeling.
\end{abstract}


\maketitle

\section{Introduction}
Quantum machine learning (QML) aims to leverage quantum mechanics to enhance learning and generative modelling~\cite{schuld2014quest, biamonte2017quantum, dunjko2018machine, ciliberto2018quantum, cerezo2022challenges}. Despite substantial progress (e.g., quantum kernels~\cite{schuld2019quantum} and variational quantum circuits~\cite{PhysRevA.98.032309, benedetti2019parameterized, cerezo2021variational}), the practical advantage over classical machine learning remains unsettled, largely due to trainability bottlenecks. In particular, optimization landscapes often exhibit plateau phenomena where gradients become extremely small; this occurs in both classical deep learning (vanishing gradients) and quantum settings (``barren plateaus'')~\cite{mcclean2018barren, ortiz2021entanglement}.

Quantum Boltzmann machines (QBMs) generalize classical Boltzmann machines by allowing non-commuting Hamiltonians~\cite{amin2018quantum}. Most existing works train such models with gradient-based methods. While recent results show that fully-visible QBMs (without hidden units) can be trained sample-efficiently~\cite{Coopmans2024}, the absence of hidden layers limits expressivity and practical modelling power.

To address this, we consider semi-quantum Boltzmann machines (sqBMs).
An sqBM is a hybrid architecture where the visible layer remains classical (commuting), while the hidden layer introduces non-commuting quantum terms. 
The restricted case of sqBMs has been proposed in \cite{Demidik2025}.
This design offers two key benefits: (i) enhanced expressive power compared to fully classical RBMs, and (ii) efficient classical simulation during training. Importantly, the lack of entanglement between visible and hidden units avoids the entanglement-induced barren plateau~\cite{ortiz2021entanglement} that plagues fully entangled QBMs~\cite{amin2018quantum}. Furthermore, this simpler structure enables closed-form expressions for parameter updates in certain algorithms.

\begin{figure*}[t]%[htbp]
  \centering
\includegraphics[width=\linewidth]{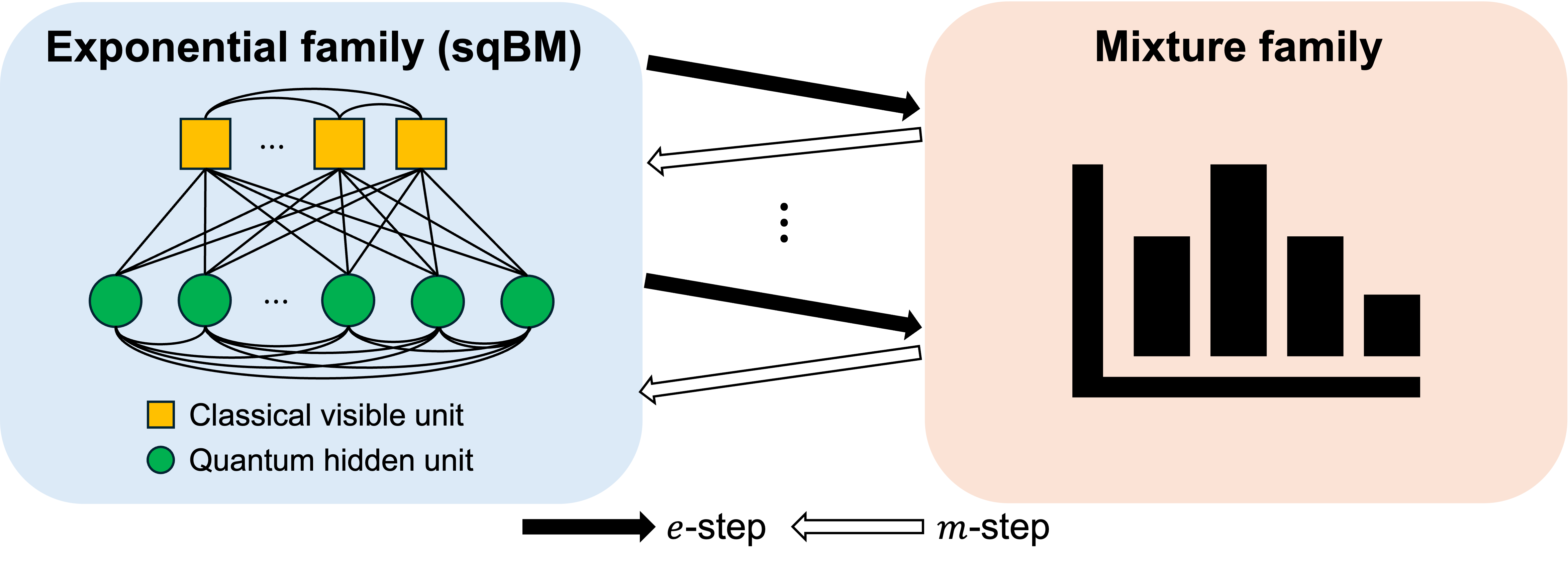}
  \caption{{\bf Conceptual overview of our learning framework.}  
This figure illustrates the em algorithm, an information-geometric generalization of the EM algorithm, proposed as an alternative approach to training QBMs. The algorithm iteratively minimizes the divergence between an exponential family (model manifold) and a mixture family (data manifold) via alternating projections. For demonstration, we apply the method to sqRBMs, where quantum effects are confined to the hidden layer.}
  \label{fig:summary}
\end{figure*}

\subsection{Classical Machine Learning Lessons: The Need for Structural Optimization.}
Before detailing QBMs, we draw crucial lessons from classical machine learning regarding the training of models with hidden variables.

To handle complex information, it is necessary to use models with hidden layers. In fact, many papers point out that models without hidden layers lack expressive power. Indeed, most current machine learning is deep learning, and many products are branded as "deep..." where "deep" refers to a deep hierarchy of hidden layers \cite{R3,R4,R5,R6}.

When using gradient methods to determine parameters in such models with hidden layers, a vanishing gradient problem occurs, preventing accurate parameter determination. This is because the complexity of the model shape due to hidden layers results in a non-convex cost function \cite{R7,R8}. In quantum machine learning, the barren plateau problem frequently arises from this gradient vanishing issue.

To solve this problem, the current practice is to decompose the target optimization problem into easier optimization subproblems that reflect the hidden layer structure. That is, a fundamental change in the optimization algorithm is necessary, rather than minor improvements like refining the gradient calculation. Many of these approaches are considered variations of the standard EM algorithm for cases with incomplete data \cite{R9,R10,R11}. Although the EM algorithm and its variants can effectively avoid the vanishing gradient problem, making it relatively easy to escape saddle points, the risk of converging to a local optimum is generally not zero in the classical case. Particularly in hidden variable models, the EM algorithm and its variants are empirically known to have a higher probability of finding better local optima compared to gradient methods, and practical performance is significantly improved \cite{R12,R13,R14}. The above represents the standard consensus in the current (non-quantum) machine learning field. We note that discussions such as Gibbs sampling pertain to the calculation method of individual steps in a specific algorithm, which is a separate issue from the behavior of the global algorithm.

\subsection{The Expressiveness-Trainability Trade-off in QBMs.}
The lessons from classical optimization highlight a critical expressiveness-versus-trainability trade-off in current QBM studies:
\begin{itemize}
    \item Coopmans \& Benedetti \cite{Coopmans2024}: This study dealt with fully visible Quantum Boltzmann machines. Since this model has no hidden layers, its expressive power is considered limited. However, the cost function is convex, meaning the Barren Plateau problem does not occur. They demonstrated that combining stochastic gradient descent with shadow tomography ensures convergence with a polynomial number of Gibbs state preparations for this model.
    \item Demidik et al. \cite{Demidik2025}: This work resolved the lack of expressive power by introducing semi quantum restricted Boltzmann machines with hidden layers. However, the algorithm used is a gradient method, which suffers from the Barren Plateau problem associated with vanishing gradients because the cost function is non-convex.
\end{itemize}
In summary, \cite{Coopmans2024} suffers from limited model expressiveness, and \cite{Demidik2025} suffers from convergence issues due to gradient vanishing. The primary challenge is to achieve high expressiveness (via hidden layers) without sacrificing efficient, guaranteed convergence (via a structural algorithm), especially when accounting for the full end-to-end computational complexity, including Gibbs state sampling.

\subsection{Our Contribution: Quantum $\text{em}$ algorithm over sqBMs.}
We propose a structured learning framework for QBMs based on a $\text{quantum em algorithm}$, an information-geometric generalization of the classical EM algorithm~\cite{dempster1977maximum, amari1992information}. Our approach leverages alternating projections between mixture and exponential families to exploit the hidden-layer structure during optimization~\cite{amari2016information, amari1995information, fujimoto2007modified, allassonniere2019new}. We extend this framework to the non-commutative (quantum) setting by leveraging recent advances in quantum information geometry \cite{hayashi2023bregman, hayashi2024reverse, hayashi2024iterative, hayashi2024generalized, hayashi2024bregman}.

While \cite{hayashi2023bregman} presented the general, highly abstract theory of the $\text{em}$ algorithm and its application to rate distortion theory, it did not provide details on the concrete calculation methods required beyond that specific example. Our paper is the first to construct the $\text{em}$ algorithm for the concrete model of the semi-quantum Boltzmann machine and demonstrate its implementation feasibility and computational efficiency.

\subsection{Method Overview and Theoretical Footing.}
As illustrated in Fig. \ref{fig:summary}, the $\text{em}$ algorithm overcomes the obstacles (lack of well-defined conditional states and non-commutativity) that hinder a direct EM extension to QBMs. Specializing to sqBMs—where the visible subspace is commuting—allows the $\text{e}$-step to become a trivial computation, significantly reducing computational cost. We formulate the $\text{m}$-step as a convex optimization problem. 
This part requires exact computation and sampling of the Gibbs state, which is a bottleneck in both quantum and classical systems.

To tackle this crucial issue of end-to-end sample and computational complexity, we show that the optimisation problem in the $\text{m}$-step is mathematically isomorphic to the training process of fully visible quantum Boltzmann machines in \cite{Coopmans2024}, allowing direct application of shadow tomography and stochastic gradient descent techniques.
By utilizing their cost reduction method, the sample complexity at m-steps (the required number of Gibbs state preparations, defined in \cite{Coopmans2024}) is polynomial.
Classical simulations to prepare a single quantum Gibbs state are typically exponentially expensive in both time and space in the worst case~\cite{bravyi2022quantum} (except in certain regimes suitable for tensor networks~\cite{PRXQuantum.2.040331, PhysRevX.11.011047}), but quantum processors naturally avoid this exponential space complexity.
Regarding time complexity, efficient Gibbs state preparation is achievable for Hamiltonians with local  structures~\cite{kastoryano2016quantum} or via various methods such as quantum Metropolis sampling, related algorithmic approaches~\cite{PhysRevLett.103.220502, temme2011quantum, chowdhury2016quantum, anschuetz2019realizingquantumboltzmannmachines, Holmes2022quantumalgorithms, chen2023quantum, zhang2023dissipative, PRXQuantum.4.010305, eassa2024gibbs, ding2025efficient, capel2025quasi} and variational methods~\cite{wu2019variational, chowdhury2020variationalquantumalgorithmpreparing, liu2021solving, consiglio2024variational, huijgen2024training, bhat2025metalearninggibbsstatesmanybody}.

Our method successfully resolves the expressiveness problem in \cite{Coopmans2024} and the convergence problem in \cite{Demidik2025}, and achieves substantially improved end-to-end computational complexity by leveraging the approach in \cite{Coopmans2024}, including the sampling cost of Gibbs state. 
Overall, our framework retains quantum hidden units while avoiding an exponential overhead of Gibbs state preparation. 
This dual advantage—architectural and algorithmic—positions our sqBM + quantum $\text{em}$ scheme as a principled step forward in quantum generative modelling.

The remainder of this paper is organized as follows.
Section~\ref{sec:2} introduces the semi-quantum Boltzmann machine (sqBM) model and fixes notation.
Section~\ref{sec:3} reviews the gradient-descent (GD) baseline and the associated training objective.
Section~\ref{sec:4} presents the proposed quantum $\text{em}$ algorithm and its specialization to the sqRBM setting.
Section~\ref{sec:5} reports numerical experiments and compares $\text{em}$ with GD on benchmark datasets.
Section~\ref{sec:6} concludes with a discussion of implications, limitations, and possible extensions.
Technical details and supplementary derivations are deferred to the Appendix.

\section{Model (semi-quantum Boltzmann machine)}\label{sec:2}
A Boltzmann machine (BM)~\cite{ackley1985learning} is an energy-based probabilistic generative model defined on an undirected graph $G=(\mathcal{V},E)$.
The vertices corresponding to binary units are classified into visible and hidden layers. The visible layer, denoted by the set of units $V=\{v_1, \ldots, v_N\}$ $(v_i = \pm 1, i \in [1, N])$, represents the inputs and outputs of the observed data. The hidden layer, denoted by $H=\{h_1, \ldots, h_M\}$, represents latent features of the data.
The edges of the graph represent the interaction pattern between units, and each edge is assigned a weight indicating the strength of the coupling.

The energy function of BM is given as  
\begin{align}
    E(\mathbf{v}, \mathbf{h}) &= \sum_{i \in V} b_i v_i+ \sum_{j \in H} b_j h_i 
    +    \sum_{(i,j)\in E_{VH}} w_{ij} v_i h_i \nonumber \\ 
    &+ \sum_{(i,j)\in E_{VV}} w_{ij} v_i v_j + 
    \sum_{(i,j)\in E_{HH}}w_{ij} h_i h_j,
\end{align}
where ${\bf v}=(v_1,...,v_N)$ and ${\bf h}=(h_1,...,h_M)$ are short-hand notations, $b_i\in\mathbb{R}$ are the bias strength and $w_{ij}\in\mathbb{R}$ are the coupling strength between $v_i$ and $h_j$. 
The edge set decomposes as \(E = E_{VH} \sqcup E_{VV} \sqcup E_{HH}\), where \(E_{VH}\) collects edges between the visible and hidden units, \(E_{VV}\) those within the visible layer, and \(E_{HH}\) those within the hidden layer.
The entire state of the RBM is represented by the probability distribution:
\begin{align}
    P_{VH, \theta} (\mathbf{v}, \mathbf{h}) := e^{- E(\mathbf{v}, \mathbf{h})} / Z,\quad Z = \sum_{\mathbf{v}, \mathbf{h}} e^{- E(\mathbf{v}, \mathbf{h})} .
\end{align}
The goal of training BM is to reproduce a probability distribution on $V$ by the marginal distribution 
\[
P_{V, \theta}({\bf v}):=\sum_{{\bf h}\in H}P_{VH,\theta}(\bf v,h)
\] 
by finding suitable parameters $\theta=(b, w)$. 

A quantum Boltzmann machine (QBM)~\cite{amin2018quantum,Demidik2025} is a model defined by replacing the binary units in the BM by qubits (1/2-spin quantum systems) $\mathbb{C}^2={\rm span}\{|\pm1\rangle\}$. The energy function of BM is replaced by a local Pauli Hamiltonian on the graph that can include non-commuting interaction terms. 

In the main analysis, we use semi-quantum BM (sqBM)~\cite{Demidik2025} with non-commutative terms only in the hidden units. As in the classical case, the vertices are decomposed as $\mathcal{V}=V\sqcup H$, where $V$ refers the visible layer and $H$ refers the hidden layer. Let us denote the Pauli operators acting on the $i$-th qubit as
\begin{align}
    \sigma_i^a = \underbrace{I \otimes \cdots \otimes I}_{i-1} \otimes \sigma^a \otimes \underbrace{I \otimes \cdots \otimes I}_{N+M-i}, \quad a=X, Y, Z.
\end{align}
Then, the Hamiltonian of sqBM is given by
\begin{align}
 H =& \sum_{i \in V} b_i \sigma_i^Z 
 + \sum_{P =X,Y,Z} \sum_{j \in H} b_j^{P} \sigma_j^{P}
 + \sum_{(i,j) \in E_{VV}}  w_{ij} \,\sigma_i^Z \sigma_j^Z \notag\\
& + \sum_{P =X,Y,Z} 
 \sum_{(i,j)\in E_{VP}} w_{ij}^{P} \,\sigma_i^Z \sigma_j^{P}\notag\\
& + \sum_{(P,Q)\in\mathcal{S}} \sum_{(i,j) \in E_{PQ}} w_{ij}^{PQ} \,\sigma_i^{P} \sigma_j^{Q},
\end{align}
where $\mathcal{S} := \{(X,X),(Y,Y),(Z,Z),(X,Y),(Y,Z),(Z,X)\}$.
The sets $E_{VV}$, $E_{VP}$, and $E_{PQ}$ denote disjoint collections of edges grouped by layers and Pauli channels:
\begin{itemize}
  \item $E_{VV} \subset V \times V$ collects visible–visible edges, each coupled through $\sigma_i^Z \sigma_j^Z$;
  \item for $P \in \{X,Y,Z\}$, $E_{VP} \subset V \times H$ collects visible–hidden edges that couple $\sigma_i^Z$ on the visible unit $v_i$ to $\sigma_j^{P}$ on the hidden unit $h_j$;
  \item for $(P,Q)\in\mathcal{S}$, $E_{PQ} \subset H \times H$ collects hidden–hidden edges that couple $\sigma_i^{P}$ to $\sigma_j^{Q}$.
\end{itemize}

The quantum state of the QBM is the quantum Gibbs state
\begin{align}
    \rho_{VH, \theta} := e^{-  H} / Z,\quad Z = \Tr[e^{-  H}],
  \label{EXP}  
\end{align}
and the reduced state on the visible layer is given by 
$\rho_{V,\theta}=\Tr_{H}\rho_{VH,\theta}$.
The probability distribution on the visible units is then
\begin{align}
    P_{V, \theta} (\mathbf{v}) = \Tr [\Lambda_{\mathbf{v}} \rho_{V, \theta}],
\end{align}
where $\Lambda_{\mathbf{v}}$ is a projection operator onto the computational basis of the visible units, given as
\begin{align}
    \Lambda_{\mathbf{v}} = \ketbra{\mathbf{v}}{\mathbf{v}}\,.
\end{align}
That is, 
$\rho_{V,\theta}$ gives our 
parametric model for the visible random variable.
In addition, the sqBM can be written as follows:
\begin{align}
    \rho_{VH, \theta} &= P_{V, \theta}\times \rho_{H|V, \theta} \\
    &:=\sum_{{\bf v}}P_{V, \theta}({\bf v})\ket{\mathbf{v}}\bra{\mathbf{v}}\otimes\rho_{H|V=\mathbf{v}, \theta}\,,
\end{align}
where
\begin{align}
    \rho_{H|V=\mathbf{v}, \theta}&=\frac{\bra{\mathbf{v}}\rho_{VH, \theta}\ket{\mathbf{v}}}{\bra{\mathbf{v}}\rho_{V, \theta}\ket{\mathbf{v}}}\,.
\end{align}

A semi-quantum restricted Boltzmann machine (sqRBM) is a sqBM in which no connections exist between units within the same layer, resulting in a bipartite graph structure. 
In the restricted case, $E_{VV}=E_{PQ}=\varnothing$; hence the Hamiltonian reduces to
\begin{align}
    H =& \sum_{i \in V} b_i \sigma_i^Z + \sum_{P = X, Y, Z} 
    \Big(\sum_{j \in H} b_j^P \sigma_j^P 
+  \sum_{(i,j)\in E_{VP}} w_{ij}^{P} \,\sigma_i^Z \sigma_j^{P}\Big).
\end{align}
Here, for each $P \in \{X,Y,Z\}$, the set $E_{VP} \subset V \times H$ denotes the visible–hidden edges that couple $\sigma_i^{Z}$ on $v_i$ to $\sigma_j^{P}$ on $h_j$.
This ensures that units within each layer are conditionally independent given the state of the opposite layer. 
The essential feature of sqRBM is its hybrid design which enables analytical derivation of output probabilities and gradients, and improves classical simulatability. Although it does not provide fully quantum enhancement, it has been theoretically shown that an RBM requires three times as many hidden units as an sqRBM to represent the same probability distribution~\cite{Demidik2025}.

\section{The gradient descent}\label{sec:3}
The cost function of BM is the Kullback–Leibler (KL) divergence $D_{\mathrm{KL}}(P_V\|P_{V, \theta}) $ defined as
\begin{align}
    &D_{\mathrm{KL}}(P_V\|P_{V, \theta}) \nonumber \\ &:= \sum_{\mathbf{v}} P_V (\mathbf{v})\left( \log P_V (\mathbf{v}) -  \log P_{V, \theta} (\mathbf{v})\right). \label{eq:KL_bm}
\end{align}
$D_{\mathrm{KL}}(P_V\|P_{V, \theta}) =0$ if and only if $P_V=P_{V,\theta}$. 
That is,
when $P_V$ is the observed empirical distribution,
the minimization of $D_{\mathrm{KL}}(P_V\|P_{V, \theta})$
is equivalent to the maximization
of the likelihood under our parametric model.
The maximum likelihood estimator 
is the most common method for finding a suitable parameter 
$\theta$, and
is given as the minimization of 
the KL-divergence.
That is, the first term of the KL-divergence is independent of $\theta$, so this cost function is equivalent to the log-likelihood
\[
Q(P_V\|P_{V, \theta})=-\sum_{\mathbf{v}} P_V (\mathbf{v})\log P_{V, \theta} (\mathbf{v})\,.
\]

The most typical method is 
the gradient descent (GD) algorithm for 
this cost function.
In this case, the parameters are updated according to the gradient of the cost function
\begin{align}
    \theta \gets \theta - \eta\cdot \partial_\theta D_{\mathrm{KL}},
    \label{VB1}
\end{align}
where $\eta>0$ is the learning rate that controls the step size. 

The gradient of BM is calculated as follows:
\begin{align}
    \partial_{\theta} D_{\mathrm{KL}} &= - \sum_{\mathbf{v}} P_V (\mathbf{v}) \sum_{\mathbf{h}} P_{H|V, \theta} (\mathbf{h}|\mathbf{v}) \partial_\theta E(\mathbf{v}, \mathbf{h}) \nonumber \\ 
    &\qquad+ \sum_{\mathbf{v}, \mathbf{h}} P_{VH, \theta} (\mathbf{v}, \mathbf{h}) \partial_\theta E(\mathbf{v}, \mathbf{h}) 
    \notag\\
    &= - (\overline{\langle \partial_\theta E(\mathbf{v}, \mathbf{h}) \rangle_{\mathbf{v}, \theta}} - \langle \partial_\theta E(\mathbf{v}, \mathbf{h}) \rangle_{\theta})
\end{align}
where
\begin{align}
    P_{H|V, \theta} (\mathbf{h}|\mathbf{v}) &= e^{-  E(\mathbf{v}, \mathbf{h})} / \sum_{\mathbf{h}} e^{-  E(\mathbf{v}, \mathbf{h})},
\end{align}
$\langle \cdot \rangle_{\theta}$ and $\langle \cdot \rangle_{\mathbf{v}, \theta}$ represent the average by $P_{VH,\theta}$ with free and fixed visible variables, respectively, and $\overline{\langle \cdot \rangle_{\mathbf{v}, \theta}} \equiv \sum_{\mathbf{v}} P_V (\mathbf{v}) \langle \cdot \rangle_{\mathbf{v}, \theta}$ represents average by the data distribution.

The GD algorithm can be applied to QBM too, where the cost function is 
given by the quantum KL divergence (quantum relative entropy) defined as
\begin{equation}
    D_{KL}(\rho\|\sigma):=\Tr\rho(\log\rho-\log\sigma)\geq0\,.
\end{equation} The gradient is then calculated as follows:
\begin{align}
    \partial_{\theta} D_{\mathrm{KL}} = - \sum_{\mathbf{v}} P_V({\bf v}) \left( \frac{\Tr [\Lambda_{\mathbf{v}} \partial_{\theta} e^{- H}]}{\Tr [\Lambda_{\mathbf{v}} e^{- H}]} - \frac{\Tr [\partial_{\theta} e^{- H}]}{\Tr [e^{- H}]} \right)\,. \label{eq:grad_qbm}
\end{align}

In QBMs, however, the presence of transverse fields leads to non-commutativity between the Hamiltonian $H$ and its parameter derivatives $\partial_\theta H$. This non-commutativity makes it difficult to express the gradient by the expectation values of local operators, as is possible in classical BMs. To address this issue, previous works such as Refs.~\cite{amin2018quantum, kieferova2017tomography} have proposed using the Golden–Thompson inequality to obtain an upper bound of the KL divergence ${\tilde D_{\mathrm{KL}}}$, thereby replacing the original optimization objective with a more tractable approximation. For this new cost function, the gradient can be written as
\begin{align}
    \partial_{\theta} {\tilde D_{\mathrm{KL}}} &= \sum_{\mathbf{v}} P_V({\bf v}) \left( \frac{\Tr [e^{- H_{\mathbf{v}}} \partial_{\theta} H_{\mathbf{v}}]}{\Tr [e^{- H_{\mathbf{v}}}]} - \frac{\Tr [e^{- H} \partial_{\theta} H]}{\Tr [e^{- H}]} \right) \notag \\
    &= - (\overline{\langle \partial_\theta H \rangle_{\mathbf{v}, \theta}} - \langle \partial_\theta H \rangle_{\theta})\,,\label{eq:gdgradient}
\end{align}
where
\begin{align}
    H_{\mathbf{v}} = \bra{\mathbf{v}} H \ket{\mathbf{v}}
\end{align}
is called the clamped Hamiltonian. In particular, for sqRBMs, the clamped Hamiltonian coincides with the model Hamiltonian, and \( D_{\mathrm{KL}} = \tilde{D}_{\mathrm{KL}} \), meaning that Eq.~\eqref{eq:gdgradient} yields the exact update rule.

In addition, various gradient-based learning algorithms have been studied for the fully visible (i.e., no hidden layer) QBM~\cite{kappen2020learning, Coopmans2024, patel2024quantum, tuysuz2024learning, patel2024natural, minervini2025evolved}, as well as for models that include hidden layers~\cite{wiebe2019generative, zoufal2021variational}. 

GD algorithms are simple to implement and widely used, but they face fundamental difficulties when applied to highly non-convex objective functions (except the fully visible case~\cite{Coopmans2024}). In such settings, convergence guarantees are generally weak, and the optimization process can be hindered by complex landscape features. In particular, GD may stagnate in regions where the objective function is nearly flat and the gradient becomes very small, even if the point is not a local minimum. This issue is well-known in e.g., variational quantum algorithms such as VQE~\cite{mcclean2018barren}, where barren plateaus significantly impair trainability. 

\section{The em algorithm}\label{sec:4}
However, GD does not give the global minimum in general
when the objective  function is not convex.
That is, there is a risk that 
the convergent of GD is trapped by a local minimum.
In the classical case, 
to reduce this risk,
an alternative to gradient-based training is the Expectation-Maximization (EM) algorithm~\cite{dempster1977maximum} that leverages the model's internal structure. 
This algorithm optimizes the model by alternating between two steps.
The E-step computes the expectation of the log-likelihood of the complete data with respect to the conditional distribution of the hidden variables, given the current parameters. The M-step then updates the model parameters by maximizing this expected log-likelihood. 

Unlike gradient-based methods, the E-step for BM is easy to implement and the M-step is a convex optimization problem, which provides theoretical guarantees of convergence and helps avoid issues such as vanishing gradients away from a local minimum. This motivates us generalization of EM to the quantum domain to mitigate the barren plateau problem in certain gradient quantum learning settings~\cite{ortiz2021entanglement}.

However, extending the classical EM algorithm to QBMs is challenging due to the absence of well-defined conditional distributions and the non-commutativity of operators. Nevertheless, the EM algorithm can be interpreted as an iterative projection method in information geometry, leading to a generalization known as the em algorithm~\cite{amari1992information}. 

To overcome the challenges of extending EM to QBMs, we consider applying a quantum version of the em algorithm, which is given by applying alternating projections for general exponential and mixture families~\cite{hayashi2023bregman}, to QBM by promoting its information-geometric formulation to the non-commutative setting.
As a concrete demonstration, we implement this algorithm in an sqRBM, where the visible layer remains commutative. This structural feature allows us to express the update rule analytically and efficiently, while still yielding a learning dynamics that differs fundamentally from both gradient-based quantum training and classical EM algorithms.

For any probability distribution $P_V$ and $\rho_{VH}$, let us denote
\begin{align}
P_V\times \rho_{H|V}&:=\sum_{{\bf v}}P_V({\bf v})\ket{\mathbf{v}}\bra{\mathbf{v}}\otimes\rho_{H|V=\mathbf{v}}\,,
\end{align}
where
\begin{align}
\rho_{H|V=\mathbf{v}}&=\frac{\bra{\mathbf{v}}\rho_{VH}\ket{\mathbf{v}}}{\bra{\mathbf{v}}\rho_{V}\ket{\mathbf{v}}}\,.
\end{align}
Then, consider the objective KL divergence:
\begin{align}
    &D_{\mathrm{KL}}(P_V\times \rho_{H|V} \| \rho_{VH, \theta}) \nonumber\\
    &= \sum_{\mathbf{v}} P_V (\mathbf{v}) 
    D_{\mathrm{KL}}(\rho_{H|V=\mathbf{v}} \| \rho_{H|V=\mathbf{v}, \theta}) \nonumber\\
    &\hspace{2cm}+ D_{\mathrm{KL}}(P_V \| P_{V, \theta}).\label{eq:klsqrbm}
\end{align}
Since the choice $\rho_{H|V=\mathbf{v}}= \rho_{H|V=\mathbf{v}, \theta}
$ vanishes the first term in the relation \eqref{eq:klsqrbm}, 
we have
\begin{align}
\min_{ \rho_{H|V} } 
D_{\mathrm{KL}}(P_V\times \rho_{H|V} \| \rho_{VH, \theta}) 
    =   
D_{\mathrm{KL}}(P_V\|P_{V, \theta}) .
\end{align}
That is, the minimization of the KL-divergence on the visible variable
can be written as the following minimization 
for $\rho_{H|V}$ and $\theta $ that can be handled by
the above em algorithm:
\begin{align}
\min_{ \theta } 
D_{\mathrm{KL}}(P_V\|P_{V, \theta}) 
=
\min_{ \rho_{H|V},\theta } 
D_{\mathrm{KL}}(P_V\times \rho_{H|V} \| \rho_{VH, \theta}) .
\end{align}
Therefore, instead of the minimization of the left hand side, 
we study the minimization of the right hand side
with respect to two variables 
$\rho_{H|V},\theta$.
To tackle this minimization, we employ the em algorithm whose detail will be explained blow.

The em algorithm is an iterative algorithm that minimizes the KL divergence between an exponential family $\mathcal{E}$ and a mixture family $\mathcal{M}$~\cite{amari1995information}.
An exponential family $\mathcal{E}$ for a random variable $X=\{x\}$ is a set of probability distributions $p(x; \boldsymbol{\theta})$ given by the exponential form:
\begin{align}
    p(X; \boldsymbol{\theta}) = \exp\left\{ \sum_{i=1}^n \theta_i r_i (x) + k(x) - \psi(\mathbf{\theta}) \right\},
\end{align}
where $\boldsymbol{\theta} = (\theta_1, \ldots, \theta_n)\in\mathbb{R}^n$ is an $n$-dimensional vector parameter, $\{r_i(x)\}_{i=1}^n, k(x)$ are functions of $x$ and $\psi$ is a normalization factor.
A mixture family $\mathcal{M}$ is a set of distributions $q(x)$ formed by a probability mixture of $m$ component distributions $\{q_i (x)\}_{i=1}^m$:
\begin{align}
    q(x) = \sum_{i=1}^m w_i q_i (x),
\end{align}
where \(\sum_{i=1}^m w_i = 1, \quad w_i \geq 0\). Typically, it is given by a set of linear constraint on expectation values.

The algorithm aims to calculate the following 
minimization of the (classical or quantum) KL divergence:
\begin{align}
    \min_{P \in \mathcal{M}, Q \in \mathcal{E}} D_{\mathrm{KL}}(P\|Q).
\end{align}
In the em algorithm, we apply the $e$-projection ($e$-step) and the $m$-projection ($m$-step) alternately.
The $e$-projection is a projection of $Q_t$ to $\mathcal{M}$ along an exponential family to find a point $P$ on a mixture family $\mathcal{M}$ that minimizes the divergence $D_{\mathrm{KL}}(P\|Q)$ as follows:
\begin{align}
    P_t = \argmin_{P \in \mathcal{M}} D_{\mathrm{KL}}(P\|Q_t).
\end{align}
This corresponds to the $E$ step in the EM algorithm.
The $m$-projection is a projection $P_t$ to $\mathcal{E}$ along a mixture family to find a point $Q$ on an exponential family $\mathcal{E}$ that minimizes the divergence $D_{\mathrm{KL}}(P\|Q)$ as follows:
\begin{align}
    Q_{t+1} = \argmin_{Q \in \mathcal{E}} D_{\mathrm{KL}}(P_t\|Q).
\end{align}
This corresponds to the $M$ step in the EM algorithm.
During this process, we have
\begin{align}
    D_{\mathrm{KL}}(P_{t-1}\|Q_t) \geq D_{\mathrm{KL}}(P_t\|Q_t) \geq D_{\mathrm{KL}}(P_t\|Q_{t+1}).
\end{align}
Generally, $e$- and $m$-projections can be written 
by using convex optimization so that it can be solved efficiently\cite{AN,hayashi2023bregman}.
The KL divergence decreases monotonically, and the algorithm converges at a local minima in the above process. 

A mixture family and an exponential family for quantum states can be straightforwardly defined \cite{AN},
and 
$\mathcal{E}=\{\rho_{VH,\theta}\mid\theta\in \mathbb{R}^{|\theta|}\}$ and
$\mathcal{M}=\{\rho_{VH}\mid \bra{\mathbf v}\rho_V\ket{\mathbf v}=P_V(\mathbf{v}),\:\forall\mathbf{v}\in V\}$
form exponential and mixture families, respectively.
Notice that any state in $\mathcal{M}$ family can be written in this form for sqBM:
\begin{equation}
    \mathcal{M}=\left\{P_V\times\rho_{H|V}\mid \rho_{VH}\right\}.
\end{equation}
The em algorithm for QBM is straightforwardly defined by quantum exponential and mixture families
\cite{hayashi2023bregman}. 
After obtaining an optimal pair $(\rho_{VH}^*,\rho_{VH,\theta}^*)\in\mathcal{M}\times \mathcal{E}$, the cost function of the QBM is bounded by
\begin{equation}
    D_{\mathrm{KL}}(P_V\|P_{V,\theta}^*)\leq D_{\mathrm{KL}}(\rho^{*}_{VH}\|\rho^*_{VH,\theta})
\end{equation}
by the data-processing inequality~\cite{lindblad1975completely}, and often the equality is achieved.
Then, we 
set the initial state to be $\rho_{VH,\theta^{(0)}}\in\mathcal{E}$.

For a given $t\in\{0,1,...\}$, each step of the em algorithm for sqBM is as follows:
\\{\bf $e$-step}: From a given state $\rho_{VH, \theta^{(t)}}\in\mathcal{E}$,  we need to find the optimal element from the mixture family $\mathcal{M}$. Since the second term in Eq.~\eqref{eq:klsqrbm} is constant over $\mathcal{M}$, the minimization is reduced to that of the first term only. Thus, given $\theta^{(t)}$, we choose
\begin{align}
    \rho_{H|V}^{(t)} &:= \argmin_{\rho_{H|V}} \sum_{\mathbf{v}} P_V (\mathbf{v}) D_{\mathrm{KL}}(\rho_{H|V=\mathbf{v}} \| \rho_{H|V=\mathbf{v}, \theta^{(t)}}).
\end{align}
Since $D_{KL}\geq0$, one can easily see that the minimum is achieved when
\begin{align}
    \rho_{H|V}^{(t)} &= \rho_{H|V, \theta^{(t)}}.
\end{align}
Note that in the case where both the visible and hidden layers are quantum, the e-step becomes nontrivial due to the absence of well-defined conditional states; see~\cite{hayashi2023bregman} for further discussion. However, in the case of sqRBM, the commutativity of the visible subspace enables a straightforward implementation of the e-step, and the relative entropy can be expressed in a tractable form.

\noindent{\bf $m$-step}: From given $P_V\times \rho_{H|V,\theta^{(t)}}$, we find the optimal element from the exponential family $\mathcal{E}$. In other words, we calculate
\begin{align}
    \theta^{(t+1)} &= \argmin_{\theta} 
    D_{\mathrm{KL}}(P_V\times\rho_{H|V, \theta^{(t)}}  \| \rho_{VH, \theta})\label{eq:thetat+1}\\
    &=\argmin_\theta \left(Z + \mathrm{Tr} [(P_V\times\rho_{H|V, \theta^{(t)}} ) H]\right).
\end{align}
Recall that $H$ is linear in $\theta=(b,w)$, and $Z=\Tr[e^{-H}]$ is a convex function of $\theta$. Therefore, this is a convex optimization in terms of $\theta$. 
Therefore a simple gradient descent method guarantees the convergence of the $m$-step, for which the update rules of the parameters are given as follows:
    
\begin{align}
    \theta &\gets \theta - \eta 
    \partial_{\theta}
    D_{\mathrm{KL}}(P_V\times\rho_{H|V, \theta^{(t)}}  \| \rho_{VH, \theta}) \label{VB2}
\\
&= \theta + \eta \mathrm{Tr}_V P_V \left( \Tr_H [\rho_{H|V, \theta^{(t)}} \partial_{\theta} H] - \Tr [\rho_{VH, \theta} \partial_{\theta} H] \right) \notag\\
    &= \theta + \eta (\overline{\langle \partial_{\theta} H \rangle_{\mathbf{v}, \theta^{(t)}}} - \langle \partial_{\theta} H \rangle_{\theta}),\label{eq:emgradient}
\end{align}
where
\begin{align}
    \overline{\langle \cdots \rangle_{\mathbf{v}, \theta}} &= \mathrm{Tr}_V P_V \Tr_H [\rho_{H|V, \theta} \cdots], \\
    \langle \cdots \rangle_{\theta} &= \Tr[\rho_{VH, \theta} \cdots].
\end{align}
See Section~\ref{appx:m-step} in Methods
for the details of the calculation. Since Eq.~\eqref{eq:thetat+1} can be regarded as a fully visible QBM on $VH$ with $P_V\times \rho_{H|V,\theta^{(t)}}$ as the target state, this step is the GD training of the fully-visible QBM in Ref.~\cite{Coopmans2024}, that shows the strict convexity and $L$-smoothness of the function ~\eqref{eq:thetat+1} (Lemmas 6, 7 in Ref.~\cite{Coopmans2024}). 
This shows that the method of~\cite{Coopmans2024} can be used, and that the m-step converges with a polynomial number of Gibbs state preparations.

While we focus on sqBMs to facilitate analytical treatment and numerical implementation, the em algorithm is not limited to this semiquantum setting. By leveraging the quantum em algorithm proposed in Ref.~\cite{hayashi2023bregman}, it can in principle be extended to general quantum Boltzmann machines with general non-commuting Hamiltonian.

The algorithm for sqBM is summarized in \cref{algo:em_qbm}.
\begin{algorithm}[H]
    \caption{The em algorithm for sqBM}
    \label{algo:em_qbm}
    \begin{algorithmic}[1]
        \AlgInput{Initial value of parameters $\theta^{(0)}$}
        \AlgOutput{Parameters $\theta$}
        \State{$\theta = \theta^{(0)}$}
        \For{$t = 0, 1, ...$}
            \State{$e$-step:
            \begin{align}
                \rho_{H|V}^{(t)} &= \rho_{H|V, \theta^{(t)}} \nonumber
            \end{align}}
            \State{$m$-step:
            \begin{align}
                &\theta^{(t+1)} = \argmin_{\theta} D_{\mathrm{KL}}(P_V\times\rho_{H|V, \theta^{(t)}}  \| \rho_{VH, \theta}) \nonumber \\
                &\theta = \theta^{(t+1)} \nonumber
            \end{align}}
            \State{End if convergence conditions are met}
        \EndFor{}
    \end{algorithmic}
\end{algorithm}

One may notice the similarity between Eq.~\eqref{eq:gdgradient} and Eq.~\eqref{eq:emgradient}. 
If the parameters are not updated until convergence but rounded up once in $m$-step, the parameter updates for the entire $e$- and $m$-steps are as follows:
\begin{align}
    \theta \gets \theta + \eta (\overline{\langle \partial_{\theta} H \rangle_{\mathbf{v}, \theta}} - \langle \partial_{\theta} H \rangle_{\theta}).\label{VB3}
\end{align}
These are exactly the update rules of the GD algorithm~\eqref{eq:gdgradient} as follows.
Therefore, the GD algorithm can be regarded as a truncated em-algorithm in this setting.

To see the above fact, 
we notice that
$D_{\mathrm{KL}}(P_V\times (\partial_\theta \rho_{H|V,\theta}) \| \rho_{VH, \theta})=0$ because the minimum 
$\min_{ \rho_{H|V} } 
D_{\mathrm{KL}}(P_V\times \rho_{H|V} \| \rho_{VH, \theta}) $
is realized with 
the choice $\rho_{H|V=\mathbf{v}}= \rho_{H|V=\mathbf{v}, \theta}
$.
Since $D_{\mathrm{KL}}(P_V\|P_{V, \theta}) 
=
D_{\mathrm{KL}}(P_V\times \rho_{H|V,\theta} \| \rho_{VH, \theta}) $,
we have
\begin{align}
&\partial_\theta D_{\mathrm{KL}}(P_V\|P_{V, \theta})
|_{\theta=\theta^{(t)}} \notag\\
=&
\partial_\theta D_{\mathrm{KL}}(P_V\times \rho_{H|V,\theta} \| \rho_{VH, \theta}) |_{\theta=\theta^{(t)}}\notag\\
=&
D_{\mathrm{KL}}(P_V\times (\partial_\theta \rho_{H|V,\theta}) \| \rho_{VH, \theta}) |_{\theta=\theta^{(t)}}\notag\\
&+
D_{\mathrm{KL}}(P_V\times \rho_{H|V,\theta} \| 
\partial_\theta \rho_{VH, \theta})|_{\theta=\theta^{(t)}} \notag\\
=&
D_{\mathrm{KL}}(P_V\times \rho_{H|V,\theta} \| 
\partial_\theta \rho_{VH, \theta})|_{\theta=\theta^{(t)}} \notag\\
=&
\partial_\theta
D_{\mathrm{KL}}(P_V\times \rho_{H|V,\theta^{(t)}} \| 
 \rho_{VH, \theta})|_{\theta=\theta^{(t)}} .
\end{align}
Therefore, the updating rule based on \eqref{VB1}
is equivalent to that on \eqref{VB2}, i.e., \eqref{VB3}.

In both algorithms, training of BM entails an inherent computational difficulty: the gradient of the KL divergence is given by the difference between two expectations, one over the data distribution (positive phase) and the other over the model distribution (negative phase). While the positive phase can often be efficiently computed, evaluating the negative phase requires sampling from the model distribution (Gibbs sampling), which involves the intractable exact calculation of partition function and becomes computationally prohibitive for large-scale networks. 

A commonly used approximation is the contrastive divergence (CD) method \cite{hinton2002training}, which estimates the negative phase by running a finite-step Markov chain Monte Carlo (MCMC) initialized from the data distribution. For EM algorithm, one can approximate the e-step by a MCMC and the m-step by a few steps of gradient descent, the overall update procedure effectively a slight generalization of a GD algorithm with a CD method, i.e.,  such algorithm is a limiting case with truncated inference and partial maximization in EM algorithm~\cite{song2016learning}. 

Although CD often performs well in practice, it does not guarantee convergence to the true local optimum. 

\begin{figure*}[htbp]%[htbp]
  % \centering
\includegraphics[width=\linewidth]{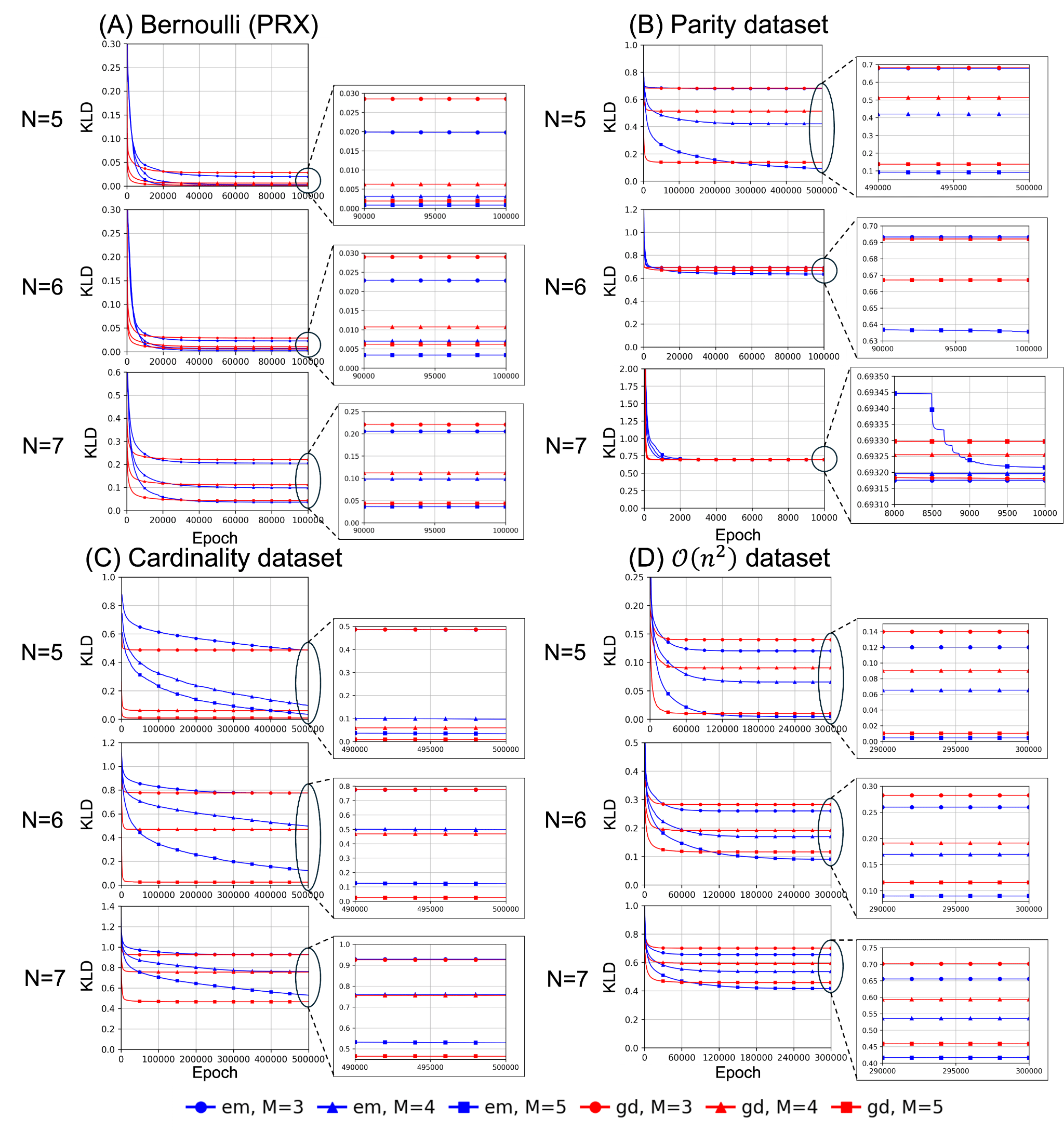}
  \caption{{\bf Performance comparison between the em algorithm and the GD method.} We train each model for a fixed number of epochs, where one epoch corresponds to a complete pass through the training data. The plot shows the final KL divergence values achieved by each algorithm on four different datasets (A, B, C, and D). Each point represents the average over 100 independent training runs. The em algorithm outperforms GD on datasets A, B, and D, while GD yields better results on dataset C. 
  }
  \label{fig:results_emgd}
\end{figure*}

\section{Experiment}\label{sec:5}
In this section, we demonstrate our em algorithm for our sqRBM to compare with the basic GD algorithm. We exploit four different data distributions: 
\begin{itemize}
    \setlength{\itemsep}{0pt}
    \setlength{\parindent}{0pt}
    \setlength{\leftskip}{0pt}
    \renewcommand{\labelitemi}{}
    \item \textbf{(A) Bernoulli} \cite{amin2018quantum}\textbf{:} 
    The training data distribution is the uniform average of     
    $K$ Bernoulli random variables
   peaked around a randomly selected center $\mathbf{s}^k = [s_1^k, s_2^k, \ldots, s_N^k]$ with $s_i^k \in \{ \pm 1 \}$.
We choose the final data distribution by 
employing the vector $\vec{\mathbf{s}}:=(\mathbf{s}^1,\ldots, \mathbf{s}^K)$.
We introduce the conditional distribution
$P_{V|S}(\mathbf{v}|\mathbf{s})
= p^{N-d(\mathbf{v},\mathbf{s})} 
(1-p)^{d(\mathbf{v},\mathbf{s})}$, where $d(\mathbf{v},\mathbf{s})$ is 
the Hamming distance between $\mathbf{v}$ and $\mathbf{s}$, and $p$ is the probability 
of each node. 
The final data distribution is obtained as:
    \begin{align}
        P_{V|\vec{S}} (\mathbf{v}|\vec{\mathbf{s}}) 
        = \frac{1}{K} \sum_{k=1}^K P_{V|S} 
        (\mathbf{v}|\mathbf{s}^k)
    \end{align}
    In this work, we use $p=0.9$ and $K=8$. We choose $\{{\bf s}^k\}$ uniformly at random once and use the same set in all experiments for fixed $N$ (see the code for the explicit bit strings).
    \item \textbf{(B) $\boldsymbol{\mathcal{O}}{\mathbf{(\textit{n}^2)}}$ dataset} \cite{Demidik2025}\textbf{:} ``$n$-bit uniform probability distribution over randomly chosen $n^2$ bitstrings.''
    \item \textbf{(C) Cardinality dataset} \cite{Demidik2025}\textbf{:} ``$n$-bit uniform probability distribution over the bitstrings that have $n/2$ cardinality.''
    \item \textbf{(D) Parity dataset} \cite{Demidik2025}\textbf{:} ``$n$-bit uniform probability distribution over the bitstrings that have even parity.''
\end{itemize}
We now describe the parameter settings and training methodology. To reduce sensitivity to the initial paramters and ensure statistical robustness, we perform 100 independent training runs for each experiment. The set of initial parameters is sampled uniformly from the interval \([-5, 5]\) in each run. For reproducibility, the random seed is fixed to 0. The training performance is evaluated by averaging the KL divergence over the 100 runs at each training step, resulting in the averaged learning curves. We use fixed hyperparameters for all models throughout this study: learning rate $\eta = 0.2$ and convergence threshold $\epsilon = 1 \times 10^{-7}$. No model-specific hyperparameter tuning is performed to ensure a fair comparison across algorithms.

We now discuss the comparative performance of the proposed em algorithm and the standard GD method. As shown in \cref{fig:results_emgd}, the final KL divergence, used as the evaluation metric, varies across datasets. The em algorithm outperforms or matches GD on three out of four datasets (A, B, and D), while GD performs better on dataset C. These results suggest that the relative advantage of each method depends on the structure of the data. Although the em algorithm demonstrates good general performance, it tends to converge more slowly than GD, which requires further improvement.

\section{Discussion}\label{sec:6}
In this study, we apply a quantum version of the em algorithm~\cite{hayashi2023bregman} to QBMs instead of extending the classical EM algorithm to QBMs.
This approach provides a principled alternative to gradient-based methods, leveraging the dual geometry of exponential and mixture families to structure the optimization process. 

Building on this perspective, we have implemented the em algorithm in a semi-quantum restricted Boltzmann machine (sqRBM)~\cite{Demidik2025}, a hybrid architecture where quantum effects appear only in the hidden layer. This choice is not incidental: sqRBMs offer enhanced expressive power compared to fully classical RBMs while remaining classically simulatable. Two structural properties make them particularly suitable for our framework. First, the absence of entanglement between visible and hidden units avoids the entanglement-induced barren plateau~\cite{ortiz2021entanglement} that plagues fully entangled QBMs~\cite{amin2018quantum}. Second, this simplified structure enables closed-form expressions for the em update rules, allowing efficient computation of each iterative step on classical hardware. These features demonstrate that our method is not a mere adaptation of an existing algorithm but a deliberate integration of architecture and learning strategy.

Crucially, our proposed sqRBM $+$ em algorithm represents an advancement over existing methods from both the model architecture and the learning algorithm perspectives. Unlike standard sqRBMs~\cite{Demidik2025} and entangled QBMs~\cite{amin2018quantum}, our approach directly intervenes and explicitly optimizes the structure of the quantum hidden units using an optimization technique inherently resistant to vanishing gradients. This dual advantage positions our framework to avoid barren plateaus while fully leveraging the expressive power of the quantum component during training.

Experimental results on multiple datasets indicate that the proposed method achieves effective learning performance on three out of four datasets, highlighting the potential of our framework in training QBMs under conditions where gradient-based methods often suffer from plateaus in the optimization landscape.

Despite these conceptual advantages, our experiments also reveal a practical limitation: the learning process often requires a large number of iterations to converge, particularly for certain data distributions. Addressing this slow convergence remains an important direction for future work. One promising strategy is to exploit the mathematical structure of the em algorithm's $m$-step, whose strong convexity and $L$-smoothness make it amenable to accelerated optimization techniques such as accelerated gradient descent. Another complementary direction is to develop approximate learning schemes. For example, Ref.~\cite{song2016learning} introduces an EM-like method that generalizes contrastive divergence (CD), offering a potential foundation for constructing fast, approximate variants of the quantum em algorithm proposed in this study.

Beyond improving convergence speed, enhancing the expressive power of the model is also a key challenge. While our analysis focused on a specific class of sqRBMs, future work could extend the method to more general architectures such as those discussed in Ref.~\cite{Demidik2025}. Furthermore, generalizing the algorithm to fully quantum RBMs, in which the visible units are also non-commutative, represents a significant step toward broader applicability. In this context, the quantum em algorithm proposed in Ref.~\cite{hayashi2023bregman} provides a valuable framework. By combining our em approach with such tools, we aim to establish a unified and scalable learning scheme for general QBMs capable of capturing more complex data distributions.

\begin{acknowledgments}
T.~K. acknowledges support from MEXT Q-LEAP JPMXS0120319794 and MEXT Q-LEAP JPMXS0120351339; from JST SPRING, Grant No. JPMJSP2125 and would like to thank the “THERS Make New Standards Program for the Next Generation Researchers.''
K.~K. acknowledges support from JSPS Grant-in-Aid for Early-Career Scientists, No. 22K13972; from MEXT-JSPS Grant-in-Aid for Transformative Research Areas (B), No. 24H00829; from JSPS KAKENHI Grant No. 23K17668; JSPS Bilateral Program (No. JPJSBP120249911).
M.H. was supported in part by
the General R\&D Projects of 1+1+1 CUHK-CUHK(SZ)-GDST Joint Collaboration Fund (Grant No. GRDP2025-022), the Guangdong Provincial Quantum Science Strategic Initiative (Grant No. 
GDZX2505003), 
the Shenzhen International Quantum Academy (Grant No. SIQA2025KFKT07),
and the National Natural Science Foundation of China under Grant 62171212.
\end{acknowledgments}

\section*{Data availability}
The data supporting the findings of this study are available from the first
author upon reasonable request.

\section*{Code availability}
The code for this study is available on Github: \url{https://github.com/txkimura/sqRBM_em}.

\appendix
\section{The em algorithm for the QBM}
In this section, we present a brief description of the em algorithm for a fully quantum Boltzmann machine (QBM) based on the Bregman divergence framework formulated by Hayashi~\cite{hayashi2023bregman}.

We consider the Hilbert space $\mathcal{H}$ and the set of density matrices $\mathcal{S}(\mathcal{H})$. 
An exponential family $\mathcal{E}$ is defined as the set of Gibbs states parameterized by $\theta \in \mathbb{R}^{|\theta|}$:
\begin{align}
    \mathcal{E} = \left\{\rho_{VH, \theta} = \exp\left(\sum_{j=1}^{|\theta|} \theta^j X_j - \mu(\theta)\right) \bigg| \theta \in \mathbb{R}^{d} \right\},
\end{align}
where $\mu(\theta) = \log(\mathrm{Tr} \exp(\sum_j \theta^j X_j))$ is partition function.
A mixture family $\mathcal{M}$ is defined by linear constraints on the expectation values:
\begin{align}
    \mathcal{M} = \left\{\rho_{VH} \in \mathcal{S}(\mathcal{H}) \bigg| \mathrm{Tr} X_j \rho_{VH} = a_j, \forall j \right\}.
\end{align}

The em algorithm alternates between projections onto these subfamilies to minimize the Bregman divergence (quantum relative entropy) in Algorithm 2 of Ref.~\cite{hayashi2023bregman}.

In the e-step, we calculate the projection to the mixture family, denoted as $\Gamma_{\mathcal{M}}^{(e)}$. 
This requires finding the unique element that satisfies the linear constraints $\mathrm{Tr} X_j \sigma = a_j$. 
Instead of solving this directly, we solve the equivalent convex optimization problem with respect to the dual parameters:
\begin{align}
    \tau^* &= \argmin_{\tau} \left( \mu \left(\tau\right) - \sum_j \tau^j a_j \right).
\end{align}
This optimization finds the unique element in $\mathcal{M}$ minimizing the divergence from the current estimate.

In the m-step, we project the result back to the exponential family, denoted as $\Gamma_{\mathcal{E}}^{(m)}$. 
We update the parameters to find the state in $\mathcal{E}$ that minimizes the Kullback-Leibler divergence (quantum relative entropy) from the state obtained in the m-step:
\begin{align}
    \theta^{(t+1)} &= \argmin_{\theta} D_{\mathrm{KL}}(\rho_{VH} \| \rho_{VH, \theta}) \label{eq:thetat+1B} \\
    &= \argmin_{\theta} \left( \mu(\theta) - \sum_j \theta^j \mathrm{Tr}[X_j \rho_{\mathrm{VH}}] \right).
\end{align}
This is the same as the m-step of sqBM.
As noted, the m-step involves a convex optimization problem, and methods such as those in~\cite{Coopmans2024} can be utilized to ensure convergence with a polynomial number of Gibbs state preparations.

\section{The em algorithm for the sqRBM model} 
In this section, we present the em algorithm for a particular instance of sq restricted RBMs with visible–hidden couplings only along the $Z$ direction. 
We set
$E_{VX}=E_{VY}=\varnothing$, $E_{VZ}=V\times H$, in addition to $E_{VV}=E_{PQ}=\varnothing$, 
and identify
$b_j^{Z}\!\equiv b_j$, $b_j^{X}\!\equiv \Gamma_j$, $b_j^{Y}=0$,
$w_{ij}^{Z}\!\equiv w_{ij}$, $w_{ij}^{X}=w_{ij}^{Y}=0$.
The Hamiltonian of this restricted sqRBM is
\begin{align}
    H = - \sum_{i\in V} b_i \,\sigma_i^{Z}
        - \sum_{j\in H} \bigl( b_j \,\sigma_j^{Z} + \Gamma_j \,\sigma_j^{X} \bigr)
        - \sum_{i \in V} \sum_{j \in H} w_{ij} \,\sigma_i^{Z} \sigma_j^{Z}.
\end{align}

In the $e$-step, we simply set
\begin{align}
    \rho_{H|V}^{(t)} &= \rho_{H|V, \theta^{(t)}}.
\end{align}

In the $m$-step, we perform the calculation of the gradient for each parameter in \cref{eq:emgradient}.
First, the positive phase for a fixed parameter $\theta^{(t)}$ can be calculated as:
\begin{align}
    \overline{\langle \partial_{b_i} H \rangle_{\mathbf{v}, \theta^{(t)}}} &= \overline{\langle \sigma_i^Z \rangle_{\mathbf{v}, \theta^{(t)}}}, \\
    \overline{\langle \partial_{b_j} H \rangle_{\mathbf{v}, \theta^{(t)}}} &= \overline{\langle \sigma_j^Z \rangle_{\mathbf{v}, \theta^{(t)}}}, \\
    \overline{\langle \partial_{\Gamma_j} H \rangle_{\mathbf{v}, \theta^{(t)}}} &= \overline{\langle \sigma_j^X \rangle_{\mathbf{v}, \theta^{(t)}}}, \\
    \overline{\langle \partial_{w_{ij}} H \rangle_{\mathbf{v}, \theta^{(t)}}} &= \overline{\langle \sigma_i^Z \sigma_j^Z \rangle_{\mathbf{v}, \theta^{(t)}}}.
\end{align}
For each term, one can further calculate the expectation values explicitly thanks to the bipartite nature of the state:
\begin{align}
    \overline{\langle \sigma_i^Z \rangle_{\mathbf{v}, \theta}} &= \sum_{v_i \in \{0, 1\}} P_V(v_i) v_i, \\
    \overline{\langle \sigma_j^Z \rangle_{\mathbf{v}, \theta}} &= \sum_{\mathbf{v}} P_V(\mathbf{v}) \frac{b_j^{\mathrm{eff}}(\mathbf{v})}{D_j(\mathbf{v})} \tanh D_j(\mathbf{v}), \\
    \overline{\langle \sigma_j^X \rangle_{\mathbf{v}, \theta}} &= \sum_{\mathbf{v}} P_V(\mathbf{v}) \frac{\Gamma_j}{D_j(\mathbf{v})} \tanh D_j(\mathbf{v}), \\
    \overline{\langle \sigma_i^Z \sigma_j^Z \rangle_{\mathbf{v}, \theta}} &= \sum_{\mathbf{v}} P_V(\mathbf{v}) v_i \frac{b_j^{\mathrm{eff}}(\mathbf{v})}{D_j(\mathbf{v})} \tanh D_j(\mathbf{v}),
\end{align}
where we define as follows:
\begin{align}
    b_j^{\mathrm{eff}}(\mathbf{v}) &= b_j + \sum_i w_{ij} v_i, \\ 
    D_j(\mathbf{v}) &= \sqrt{\Gamma^2_j + (b_j^{\mathrm{eff}}(\mathbf{v}))^2}.
\end{align}
This part is almost same as the GD of QBM~\cite{Demidik2025}, so we omit the further details of these calculations here. The positive phase is constant of $\theta$, so for each $t$, we only calculate it once.

The gradient appearing in the negative phase is given as:
\begin{align}
    \langle \partial_{b_i} H \rangle_{\theta} &= \langle \sigma_i^Z \rangle_{\theta}, \\
    \langle \partial_{b_j} H \rangle_{\theta} &= \langle \sigma_j^Z \rangle_{\theta}, \\
    \langle \partial_{\Gamma_j} H \rangle_{\theta} &= \langle \sigma_j^X \rangle_{\theta}, \\
    \langle \partial_{w_{ij}} H \rangle_{\theta} &= \langle \sigma_i^Z \sigma_j^Z \rangle_{\theta}.
\end{align}
This part depends on the optimizing parameter $\theta$, so we repeatedly calculate the following and update the parameters until the convergence condition is met:
\begin{align}
    P_{V, \theta}(\mathbf{v}) &= \exp(- \sum_i b_i v_i) \prod_j \cosh (D_j(\mathbf{v})) / Z, \\
    Z &= \sum_{\mathbf{v}} \exp(- \sum_i b_i v_i) \prod_j \cosh(D_j(\mathbf{v})), \\
    \langle \sigma_i^Z \rangle_{\theta} &= \sum_{v_i \in \{0, 1\}} P_{V, \theta}(v_i) v_i, \\
    \langle \sigma_j^Z \rangle_{\theta} &= \sum_{\mathbf{v}} P_{V, \theta}(\mathbf{v}) \frac{b_j^{\mathrm{eff}}(\mathbf{v})}{D_j(\mathbf{v})} \tanh D_j(\mathbf{v}), \\
    \langle \sigma_j^X \rangle_{\theta} &= \sum_{\mathbf{v}} P_{V, \theta}(\mathbf{v}) \frac{\Gamma_j}{D_j(\mathbf{v})} \tanh D_j(\mathbf{v}), \\
    \langle \sigma_i^Z \sigma_j^Z \rangle_{\theta} &= \sum_{\mathbf{v}} P_{V, \theta}(\mathbf{v}) v_i \frac{b_j^{\mathrm{eff}}(\mathbf{v})}{D_j(\mathbf{v})} \tanh D_j(\mathbf{v}).
\end{align}
The proof of the above calculations can also be found in Ref.~\cite{Demidik2025}.

The convergence condition of the $m$-step is set by the difference of the KL divergence:
\begin{align}
    \Delta \mathrm{KL} := D_{\mathrm{KL}}&(P_V\times\rho_{H|V, \theta^{(t)}}  \| \rho_{VH, \theta'}) \nonumber \\
    &- D_{\mathrm{KL}}(P_V\times\rho_{H|V, \theta^{(t)}}  \| \rho_{VH, \theta})\leq\epsilon,
\end{align}
where $\theta'$ is the next updated parameter after $\theta$.
The detailed calculations are given in Subsection~\ref{appx:qre}.

This more detailed algorithm for our sqRBM is summarized in \cref{algo:em_sqRBM}.

\begin{algorithm}[H]
    \caption{The em algorithm for our sqRBM}
    \label{algo:em_sqRBM}
    \begin{algorithmic}[1]
        \AlgInput{Initial value of parameters $\theta^{(0)} = (b_i^{(0)}, b_j^{(0)}, \Gamma_j^{(0)}, w_{ij}^{(0)})$}
        \Comment{index $i$ for visible unit, $j$ for hidden unit}
        \AlgOutput{Parameters $\theta = (b_i, b_j, \Gamma_j, w_{ij})$}
        \State{$\theta = \theta^{(0)}$}
        \For{$\mathrm{epoch} = 0, \cdots, \mathrm{n\_epochs}-1$}
            \State{$\overline{\langle \sigma_i^Z \rangle_{\mathbf{v}}} \gets \overline{\langle \sigma_i^Z \rangle_{\mathbf{v}, \theta}}$}
            \State{$\overline{\langle \sigma_j^Z \rangle_{\mathbf{v}}} \gets \overline{\langle \sigma_j^Z \rangle_{\mathbf{v}, \theta}}$}
            \State{$\overline{\langle \sigma_j^X \rangle_{\mathbf{v}}} \gets \overline{\langle \sigma_j^X \rangle_{\mathbf{v}, \theta}}$}
            \State{$\overline{\langle \sigma_i^Z \sigma_j^Z \rangle_{\mathbf{v}}} \gets \overline{\langle \sigma_i^Z \sigma_j^Z \rangle_{\mathbf{v}, \theta}}$}
            \For{$\mathrm{epoch\_m} = 0, \cdots, \mathrm{n\_epochs\_m}-1$}
                \State{$b_i \gets b_i + \eta (\overline{\langle \sigma_i^Z \rangle_{\mathbf{v}}} - \langle \sigma_i^Z \rangle_{\theta})$}
                \State{$b_j \gets b_j + \eta (\overline{\langle \sigma_j^Z \rangle_{\mathbf{v}}} - \langle \sigma_j^Z \rangle_{\theta})$}
                \State{$\Gamma_j \gets \Gamma_j + \eta (\overline{\langle \sigma_j^X \rangle_{\mathbf{v}}} - \langle \sigma_j^X \rangle_{\theta})$}
                \State{$w_{ij} \gets w_{ij} + \eta (\overline{\langle \sigma_i^Z \sigma_j^Z \rangle_{\mathbf{v}}} - \langle \sigma_i^Z \sigma_j^Z \rangle_{\theta})$}
                \State{End if $|\Delta \mathrm{QRE}| < \epsilon$}
            \EndFor{}
        \EndFor{}
    \end{algorithmic}
\end{algorithm}

% The \nocite command causes all entries in a bibliography to be printed out
% whether or not they are actually referenced in the text. This is appropriate
% for the sample file to show the different styles of references, but authors
% most likely will not want to use it.
% \nocite{*}

\if0
\clearpage

\onecolumngrid
%\appendix
%\section*{Appendix}

\makeatletter
% Redefine subsection numbering for the appendix
\renewcommand\thesubsection{\thesection\arabic{subsection}}
% Ensure subsection titles use the modified numbering
\renewcommand\p@subsection{}
\makeatother
\fi

%\subsection{Proofs}
\section{Proof of the calculation of the m-step} \label{appx:m-step}
The $m$-step is given as follows:
\begin{align}
    \theta^{(t+1)} =& \argmin_{\theta} D_{\mathrm{KL}}(P_V\times\rho_{H|V, \theta^{(t)}}  \| \rho_{VH, \theta}) \notag \\
    =& \argmin_{\theta} \Tr (P_V\times\rho_{H|V, \theta^{(t)}} )\notag \\
    & \cdot (\log (P_V\times\rho_{H|V, \theta^{(t)}} ) - \log \rho_{VH, \theta}).
\end{align}
Since $\Tr (P_V\times\rho_{H|V, \theta^{(t)}} ) (\log (P_V\times\rho_{H|V, \theta^{(t)}} ))$ does not depend on $\theta$, we minimize $- \Tr (P_V\times\rho_{H|V, \theta^{(t)}} ) \log \rho_{VH, \theta}$.
We have
\begin{align}
&    - \Tr (P_V\times\rho_{H|V, \theta^{(t)}} ) \log \rho_{VH, \theta} 
   \notag \\
    &= Z + \Tr (P_V\times\rho_{H|V, \theta^{(t)}} ) H.
\end{align}
Thus, we have
\begin{align}
    \theta^{(t+1)} := \argmin_\theta Z + \mathrm{Tr} (P_V\times\rho_{H|V, \theta^{(t)}} ) H.
\end{align}
The GD to minimize $\theta$ is given as follows: 
\begin{align}
    \theta &\gets \theta + \eta \big(\partial_{\theta} Z + \mathrm{Tr} (P_V\times\rho_{H|V, \theta^{(t)}} ) \partial_{\theta} H \big) \notag \\
    &= \theta + \eta \big( \mathrm{Tr}_V P_V \partial_{\theta} Z + \mathrm{Tr} (P_V\times\rho_{H|V, \theta^{(t)}} ) \partial_{\theta} H \big) \notag \\
    &= \theta + \eta \mathrm{Tr}_V P_V \big( \partial_{\theta} Z + \mathrm{Tr}_H \rho_{H|V, \theta^{(t)}} \partial_{\theta} H \big).
\end{align}
The derivative of $Z$ with respect to $\theta$ is given as follows:
\begin{align}
  &  \partial_{\theta} Z \notag \\
    =& \frac{1}{\mathrm{Tr} \exp (-H)} \partial_{\theta} \mathrm{Tr} \exp(-H) \notag \\
    =& \frac{1}{\mathrm{Tr} \exp (-H)} \mathrm{Tr} \partial_{\theta} \exp(-H)\notag  \\
    =& \frac{1}{\mathrm{Tr} \exp (-H)} \notag \\
   &\cdot \mathrm{Tr} \left( - \sum_{m=1}^n e^{-m \delta \tau H} \partial_{\theta} H \delta t e^{-(n-m) \delta \tau H} + \mathcal{O}(\delta \tau^2) \right) \notag \\
    &\xrightarrow[n \to \infty]{} \frac{1}{\mathrm{Tr} \exp (-H)} \mathrm{Tr} \left( - \int_0^1 d\tau e^{- \tau H} \partial_{\theta} H e^{(\tau-1) H} \right) \notag \\
    =& - \mathrm{Tr} \int_0^1 d\tau \frac{ e^{- \tau H} \partial_{\theta} H e^{(\tau-1) H}}{\mathrm{Tr} \exp{(-H)}} \notag \\
    =& - \Tr \frac{e^{-H} \partial_{\theta} H}{\Tr e^{-H}} \notag \\
    =& - \Tr \rho_{VH, \theta} \partial_{\theta} H.
\end{align}
Therefore, the GD to minimize $\theta$ is given as follows: 
\begin{align}
    \theta \gets \theta + \eta \mathrm{Tr}_V P_V \left( \Tr_H [\rho_{H|V, \theta^{(t)}} \partial_{\theta} H] - \Tr [\rho_{VH, \theta} \partial_{\theta} H] \right).
\end{align}

\section{The calculation of the difference of quantum relative entropy} \label{appx:qre}
The calculation of the difference of quantum relative entropy is given as follows:
\begin{align}
    \Delta \mathrm{QRE} =& D_{\mathrm{KL}}(P_V\times\rho_{H|V, \theta^{(t)}}  \| \rho_{VH, \theta'}) \notag \\
    &- D_{\mathrm{KL}}(P_V\times\rho_{H|V, \theta^{(t)}}  \| \rho_{VH, \theta}) \notag \\
    =& \Tr [(\rho_{H|V} \times P_V)(\log \rho_{VH, \theta'} - \log \rho_{VH, \theta})].
\end{align}
Therefore, we perform the following calculation in each step of the $m$-step:
\begin{align}
&    \Tr [(P_V\times\rho_{H|V, \theta^{(t)}} ) \log \rho_{VH, \theta}] \notag \\
    =& - \Tr [(P_V\times\rho_{H|V, \theta^{(t)}} ) H] - \Tr [(P_V\times\rho_{H|V, \theta^{(t)}} ) \log Z] \notag \\
    =& - \Tr_V P_V \Tr_H [\rho_{H|V, \theta^{(t)}} H] - \log Z \notag \\
    =& - \sum_j \Gamma_j \Tr_H [\rho_{H|V, \theta^{(t)}} \sigma_j^X] - \sum_i b_i \Tr_V [P_V \sigma_i^Z] \notag\\
    & - \sum_j b_j \Tr_H [\rho_{H|V, \theta^{(t)}} \sigma_j^X] \nonumber \\ 
    & - \sum_{i, j} w_{ij} \Tr_V P_V \Tr_H [\rho_{H|V, \theta^{(t)}} \sigma_j^X] - \log Z \notag \\
    =& - \sum_j \Gamma_j \overline{\langle \sigma_j^X \rangle_{\mathbf{v}, \theta}} - \sum_i b_i \overline{\langle \sigma_i^Z \rangle_{\mathbf{v}, \theta}} - \sum_j b_j \overline{\langle \sigma_j^Z \rangle_{\mathbf{v}, \theta}} \notag \\
  &  - \sum_{i, j} w_{ij} \overline{\langle \sigma_i^Z \sigma_j^Z \rangle_{\mathbf{v}, \theta}} - \log Z.
\end{align}

\section{Performance comparison of the em algorithm for sqRBM and RBM}
\begin{figure*}[htbp]%[htbp]
  % \centering
\includegraphics[width=\linewidth]{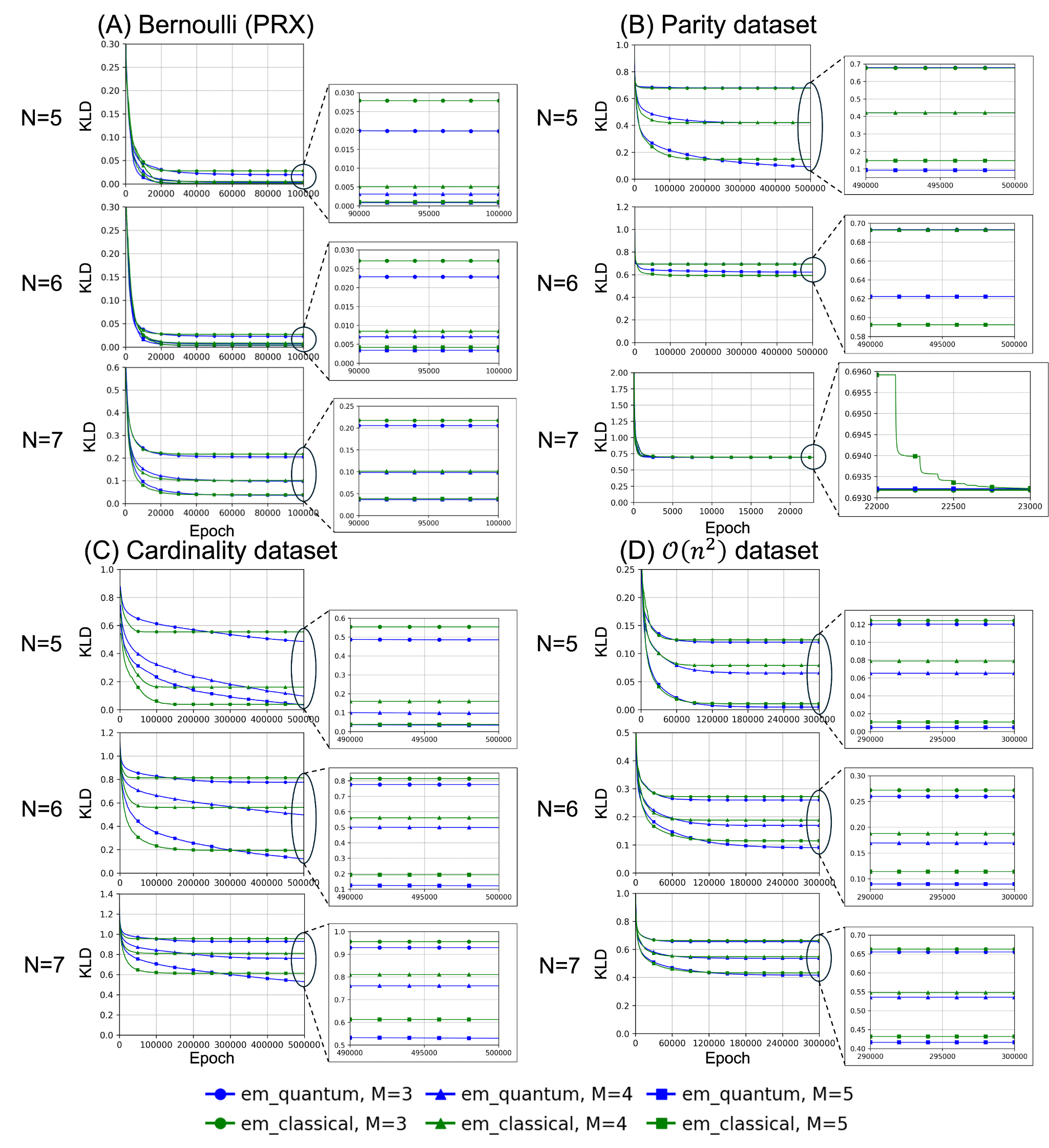}
  \caption{{\bf Performance comparison of the em algorithm for sqRBM and RBM.} 
  We train each model for a fixed number of epochs, where one epoch corresponds to a complete pass through the training data. 
  The plot shows the final KL divergence values achieved by each algorithm on four different datasets (A, B, C, and D). 
  Each point represents the average over 100 independent training runs.
  The em algorithm for sqRBM outperforms RBM except for one result. 
  }
  \label{fig:results_emem}
\end{figure*}

This section compares the performance of em algorithm for the proposed sqRBM against its application to the conventional RBM, aiming to validate the effectiveness of quantum terms.
The results of the performance comparison are presented in \cref{fig:results_emem}.

The em algorithm for sqRBM shows equal or superior performance to RBM except one result.
This can be attributed to the high expressive power that sqRBM has acquired by incorporating quantum effects.
In $N=6$ for dataset B, the RBM results outperform the sqRBM results.
One possible reason for this is that sqRBM is more complex than RBM, resulting in slower convergence, and the optimal solution may not have been reached in the set number of training cycles.

Although only a quantum term for the Pauli X component is introduced in the hidden layer in this study, adding a term for the Pauli Y component, as shown in \cite{Demidik2025}, may further improve the expressiveness of the model and improve its performance.

\bibliographystyle{apsrev4-2}
\bibliography{ref}% Produces the bibliography via BibTeX.

% (APS/PRX 投稿では、Author contributions / Competing interests は
%  投稿システム側で入力する運用が多く、本文に残すと Nature 風味が出やすいので一旦外す)
% \section*{Author contributions}
% T. K. was responsible for the main coding, numerical experiments, writing most of the manuscript, and preparing the figures. K.K. supervised the code, experiments, and manuscript revision. M. H. led the theoretical analysis, conceptualization, and proposed the idea of applying the em algorithm to the semi-quantum restricted Boltzmann machine. This work represents a collaboration among all three authors. All authors reviewed and approved the manuscript.

% \section*{Competing interests}
% The authors declare no competing interests.

\end{document}